\documentclass[aps,paper,12pt]{revtex4-1}

\usepackage{amsmath}
\usepackage{graphicx}
\usepackage{dcolumn}
\usepackage{bm}
\usepackage{epsfig}
\usepackage{graphics}
\usepackage{rotating}
\usepackage{xcolor}
\graphicspath{{./figures/}}
\usepackage{float}
\begin{document}


\title{Gravitational particle production and the Hubble tension}
\author{Recai Erdem}
\email{recaierdem@iyte.edu.tr}
\affiliation{Department of Physics \\
{\.{I}}zmir Institute of Technology\\
G{\"{u}}lbah{\c{c}}e, Urla 35430, {\.{I}}zmir, T{\"{u}}rkiye}


\begin{abstract}
The effect of gravitational particle production of scalar particles on the total effective cosmic energy density (in the era after photon decoupling till the present) is considered. The effect is significant for heavy particles. It is found that gravitational particle production results in an effective increase in the directly measured value of the Hubble constant $H_0$ while it does not affect the value of the Hubble constant in the calculation of the number density of baryons at present time that is used to calculate recombination redshift. This may explain why the Hubble constant determined by local measurements and non-local measurements (such as CMB) are different. This suggests that gravitational particle production may have a non-negligible impact on the $H_0$ tension.

\end{abstract}


\maketitle

\section{introduction}

Gravitational particle production is a generic property of quantum fields in time-dependent backgrounds such as the Friedman-Lema{\^{i}tre-   Robertson-Walker (FLRW) spacetimes \cite{QFTC1,QFTC2}. For example, the solution of the effective equation of motion of a free scalar field (namely, the mode function) at two different times, in general, is different since the equation of motion contains a time dependent effective mass. Hence, there exist different vacua at different times (that are described by different creation and annihilation operators and different mode functions). The mode function (and the corresponding creation/annihilation operators) at a given time may be expressed in terms of the mode function (and the corresponding creation/annihilation operators) at another time by a Bogolyubov transformation. Thus, a mode function at an initial time (that describes an "in" state) evolves into another value at a later time that may be expanded in terms of the mode function at that time (namely, the mode function of the "out" state).  This is the well-known gravitational particle production. Therefore, gravitational particle production is a generic process for quantum fields in FLRW spacetimes. Hence, gravitational particle production necessarily takes place in cosmology. The aim of this study is to see the degree of the impact of this process on the standard cosmology through the example of a scalar field in the era after the photon decoupling till the present.

Hubble tension is the huge discrepancy between the direct local measurements of Hubble constant by type SN Ia supernovas calibrated by Cepheids \cite{Riess} and the measurements of Planck \cite{Planck} and other non-local measurements such as baryon acoustic oscillations (BAO) \cite{BAO1,BAO} imprinted on galaxy autocorrelation functions (that also involve effects of much earlier times and assume $\Lambda$CDM). The values of Hubble constant obtained from local measurements are almost certainly higher than the ones that also include the effect of higher redshifts. For example, \cite{Riess} finds the Hubble constant as $\left(73.04\pm\,1.04\right)km\,s^{-1}\,Mpc^{-1}$ while \cite{Planck} finds it as $\left(67.4\pm\,0.5\right)km\,s^{-1}\,Mpc^{-1}$. SN Ia supernova and Planck measurements differ by at least 5$\sigma$ \cite{Riess,Planck,tension,Kamionkowski,Dainotti}. This is called Hubble tension. There are many different approaches and models proposed for solution of the Hubble tension problem \cite{tension,Kamionkowski,tension2,jump-G1,jump-G3,jump-G2,Uzan,Pan,Montani1,Montani2,Montani3,Montani4,Montani5,Montani6}. The standard approach of the theoretical models that attempt to solve this problem is to assume the value of Hubble constant obtained by local measurements to be the correct one, and to seek a model that makes the results of Planck (and other non-local measurements) compatible with local measurements. In this vein, they try to modify $\Lambda$CDM at late times (close to the present time) or early times (just before the time of recombination) or at both epochs so that the equations given below result in the same result as the local measurements. In this study a different approach is adopted. The effect of gravitational particle production of scalar particles on Hubble constant is considered. It is shown that, depending on the value of the total mass of the scalars in the model, inclusion of the effect of gravitational particle production in the context of $\Lambda$CDM may ameliorate or relieve the Hubble tension.

In the following, first, in Section II, the basic concepts and techniques necessary for a better understanding of the present study are briefly reviewed. In Section III, it is shown that  adiabatic approximation that is used in the present study is applicable to the era after photon decoupling in $\Lambda$CDM for a wide range of scalar particle masses. In Section IV, the contribution of gravitational particle production to energy density is discussed. In Section V, the implications of gravitational particle production for Hubble tension are discussed. Finally, Section VI summarizes the main conclusions.

\section{preliminiaries}
Spacetime at cosmological scales may be described by spatially flat Robertson-Walker metric
\begin{equation}
ds^2\,=\,-dt^2\,+\,a^2(t)\left[dr^2+r^2
(d\theta^2+\sin^2{\theta}d\phi^2)\right]\;. \label{e1}
\end{equation}

We consider the following action for a scalar field $\phi$ in this space
\begin{eqnarray}
&S&
\,=\,\int\sqrt{-g}\;d^4x\,\frac{1}{2}\left[-g^{\mu\nu}\partial_\mu\phi\partial_\nu\phi\,-\,m_\phi^2\phi^2\right] \label{e2a} \,=\,
\int\,d^3x\,d\eta\,\frac{1}{2}\left[
\tilde{\phi}^{\prime\;2}-(\vec{\nabla}\tilde{\phi})^2
-\tilde{m}_\phi^2\tilde{\phi}^2\right]\;,
\label{e2b}
\end{eqnarray}
where $m_\phi$ is the mass of $\phi$, prime denotes derivative with respect to conformal time $\eta$ \cite{QFTC1} (while an over-dot denotes the derivative with respect to $t$) and
\begin{equation}
d\eta\,=\,\frac{dt}{a(t)}~,
~\tilde{\phi}\,=\,a(\eta)\;\phi~,~~~
\tilde{m}_\phi^2\,=\,m_\phi^2a^2-\frac{a^{\prime\prime}}{a}.
\label{e2aa}
\end{equation}

The field $\tilde{\phi}$ may be expressed as
\begin{equation}
\tilde{\phi}(\vec{x},\eta)\,=\,\frac{1}{\sqrt{2}}\,\int\,\frac{d^3k}{(2\pi)^\frac{3}{2}}\,\left[ e^{i\vec{k}.\vec{x}}\,v_k^*(\eta)\,\hat{a}_{\vec{k}}^-
\,+\, e^{-i\vec{k}.\vec{x}}\,v_k(\eta)\,\hat{a}_{\vec{k}}^+ \right] \label{eq:3}
\end{equation}
The mode function $v_k(\eta)$ satisfies the equation of motion for $\tilde{\phi}$
\begin{equation}
v_k^{\prime\prime}\,+\,\omega_k^2\,v_k = 0
\label{eq:4}
\end{equation}
where
\begin{equation}
\omega_k\,=\,\sqrt{\vec{k}^2\,+\,\tilde{m}_\phi^2}
\label{eq:5}
\end{equation}

Vacuum state is the ground state with minimum energy. In curved space, in general, the ground state at an instant of time is not the ground state at another time. Hence, the annihilation operators corresponding to the corresponding vacuum state at a given time do not destroy the vacuum state at another time. Therefore, the annihilation and creation operators and the mode functions at different times are different in general in curved spaces \cite{QFTC1}. The field $\tilde{\phi}$ in another vacuum (other than the one specified in (\ref{eq:3})) with annihilation and creation operators $\hat{b}_{\vec{k}}^-$ and $\hat{b}_{\vec{k}}^+$ may be expanded as
\begin{equation}
\tilde{\phi}(\vec{x},\eta)\,=\,\frac{1}{\sqrt{2}}\,\int\,\frac{d^3k}{(2\pi)^\frac{3}{2}}\,\left[ e^{i\vec{k}.\vec{x}}\,u_k^*(\eta)\,\hat{b}_{\vec{k}}^-
\,+\, e^{-i\vec{k}.\vec{x}}\,u_k(\eta)\,\hat{b}_{\vec{k}}^+ \right] \label{eq:6}
\end{equation}
where $u_k$ satisfies the same equation as (\ref{eq:4}) and is related to $v_k$ and $v_k^*$ by
\begin{equation}
u_k(\eta)\,=\,\alpha_k\,v_k(\eta)\,+\,\beta_k\,v_k^*(\eta)  \label{eq:7}
\end{equation}
where $\alpha_k$, $\beta_k$ are called Bogolyubov coefficients. In similar fashion, $\hat{b}_{\vec{k}}^-$ is related to $\hat{a}_{\vec{k}}^-$ and $\hat{a}_{\vec{k}}^+$ by
\begin{equation}
\hat{b}_{\vec{k}}^-\,=\,\alpha_k\,\hat{a}_{\vec{k}}^-\,-\,\beta_k\,\hat{a}_{\vec{k}}^+ \label{eq:7}
\end{equation}

Mode functions may be expressed in a WKB-approximation-like form
\begin{equation}
v_k(\eta)\,=\,\frac{1}{\sqrt{W_k(\eta)}}\,\exp{\left[i\int_{\eta_0}^\eta\,W_k(\eta)\,d\eta \right]}  \label{eq:8}
\end{equation}
where (\ref{eq:4}) implies that $W_k(\eta)$ should satisfy
\begin{equation}
W_k^2\,=\,\omega_k^2\,-\, \frac{1}{2}\left[ \frac{W_k^{\prime\prime}}{W_k}-\frac{1}{2}\left(\frac{W_k^{\prime}}{W_k}\right)^2 \right]\,+\,\frac{i}{2}\,W_k^\prime
 \label{eq:9}
\end{equation}

The following $W_k$ that approximately satisfies (\ref{eq:9}) may be adopted as an approximate solution
\begin{equation}
W_k\,\simeq\,\omega_k~~~~\mbox{if}~~~\frac{\omega_k^\prime}{\omega_k^2}\,\ll\,1~~~~\mbox{and}~~~~\frac{\omega_k^{\prime\prime}}{\omega_k^3}\,\ll\,1.  \label{eq:10}
\end{equation}
(\ref{eq:10}) may be identified as adiabatic conditions \cite{QFTC2,Boyanovsky}.

\section{applicibality of the adiabatic conditions to the $\Lambda$CDM universe after the decoupling}

In this section we show that the adiabatic conditions (\ref{eq:10}) are satisfied in $\Lambda$CDM after the time of decoupling for a wide range of $m_\phi$. Moreover, we find that (unlike their standard form) the adiabatic conditions in this case are satisfied independent of the value of $|\vec{k}|$ (in the above mentioned intervals). To this end, first we show that $\left|\frac{\tilde{m}_\phi^\prime}{\tilde{m}_\phi^2}\right|\,\ll\,1$ and $\left|\frac{\tilde{m}_\phi^{\prime\prime}}{\tilde{m}_\phi^3}\right|\,\ll\,1$ are satisfied in $\Lambda$CDM for a wide range of $m_\phi$, and then we obtain the corresponding adiabatic conditions.

$\tilde{m}_\phi^2$ in (\ref{e2b}) may be expressed as
\begin{equation}
\tilde{m}_\phi^2\,=\,m_\phi^2a^2-\frac{a^{\prime\prime}}{a}\,=\,m_\phi^2a^2\,-\,a\left(H^\prime+2a\,H^2\right)
\,=\,m_\phi^2a^2\,-\,a^2H\left(a\,\frac{dH}{da}+2\,H\right),
\label{eq:14}
\end{equation}
where $H=\frac{\dot{a}}{a}=\frac{a^\prime}{a^2}$ is Hubble parameter and $H^\prime=a^2H\frac{dH}{da}$ is employed.
Then, we obtain
\begin{eqnarray}
&&\left(\tilde{m}_\phi^2 \right)^\prime
\,=\,
2\,m_\phi^2\,a^3\,H
\,-\,7a^4\,H^2\,\left(\frac{dH}{da}\right)\,-\,4a^3H^3\,-\,a^5H\left(\frac{dH}{da}\right)^2\,-\,a^5H^2\left(\frac{d^2H}{da^2}\right),
\label{eq:13a} \\
&&\left(\tilde{m}_\phi^2 \right)^{\prime\prime}
\,=\,a^2H\left[\frac{d\left(\tilde{m}_\phi^2\right)^\prime}{da}\right]
\,=\,a^4H{\Bigg[}6m^2H-12H^3+\left(2m^2a-40\,aH^2\right)\left(\frac{dH}{da}\right)
-19a^2H\left(\frac{dH}{da}\right)^2\nonumber \\
&&-12a^2H^2\left(\frac{d^2H}{da^2}\right)
-a^3\left(\frac{dH}{da}\right)^3-4a^3H\left(\frac{dH}{da}\right)\left(\frac{d^2H}{da^2}\right)-a^3H^2\left(\frac{d^3H}{da^3}\right)\Bigg].
\label{eq:13b}
\end{eqnarray}

The Hubble parameter for $\Lambda$CDM (that describes the background evolution) is
\begin{equation}
H\,=\,H_0\,\sqrt{\Omega_\Lambda\,+\,\Omega_M\,a^{-3}\,+\,\Omega_R\,a^{-4}}  \label{eq:15}
\end{equation}
where $\Omega_\Lambda$, $\Omega_M$, $\Omega_R$ are the density parameters for cosmological constant, dust, radiation, respectively.
(In fact, (\ref{eq:15}) is expected to} approximately hold in extensions of $\Lambda$CDM as well since $\Lambda$CDM seems to be in agreement with observations  at cosmological scales except for a few potential problems including $H_0$ tension.)
Use of (\ref{eq:15}) in (\ref{eq:14}) results in
\begin{equation}
\tilde{m}_\phi^2\,=\,m_\phi^2a^2\,\left\{1\,-\,2\left(\frac{H_0}{m_\phi}\right)^2\,\Omega_\Lambda\left[1\,
+\,\frac{1}{4}\left(\frac{\Omega_M}{\Omega_\Lambda}\right)\,a^{-3}\right] \right\}.
\label{eq:18}
\end{equation}
 We observe that the term that is proportional to $H_0^2$ in (\ref{eq:18}) is larger at smaller scale factors. Therefore, for $a(\eta)\,>\,10^{-3}$, this term has the largest value at the beginning of decoupling $a(\eta)\,\sim\,10^{-3}\,>\,10^{-4}$. Thus, for $a(\eta)\,>\,10^{-4}$ we have
 \begin{equation}
1\,+\,\frac{1}{4}\left(\frac{H_0}{m_\phi}\right)^2\Omega_M\,a^{-3}\,<\,\frac{1}{4}\left(\frac{H_0}{m_\phi}\right)^2\Omega_M\,10^{12}.
\label{eq:20}
\end{equation}
Hence, the term proportional to $H_0^2$ in (\ref{eq:18}) is negligible for scale factors greater than  $a(\eta)\,\sim\,10^{-4}$ if
\begin{equation}
\frac{1}{2}\left(\frac{H_0}{m_\phi}\right)^2\Omega_M\,10^{12}~\,\ll\,1.
\label{eq:21}
\end{equation}
This, in turn, means that
\begin{eqnarray}
&&\left(\frac{H_0\hbar}{m_\phi\,c^2}\right)^2\Omega_M\,10^{12}\,\ll\,1~~~\Rightarrow~~~~~\left(\frac{m_\phi\,c^2}{eV}\right)\,\gg\,10^{-27}
\label{eq:23}
\end{eqnarray}
where $c$ and $\hbar$ are written explicitly in (\ref{eq:21}) to obtain the left-hand side of (\ref{eq:23}) and it is multiplied and divided by $(eV)^2$ and then rearranged and $H_0\hbar\,\simeq\,1.5\times\,10^{-33}eV$ is used to obtain the right-hand side of (\ref{eq:23}).
Eq.(\ref{eq:23}) and Eq.(\ref{eq:18}) imply that
\begin{equation}
\tilde{m}_\phi^2\,\simeq\,m_\phi^2a^2~~~~~\mbox{provided that}~~~~m_\phi\,c^2\,\gg\,10^{-27}\,eV.
\label{eq:24}
\end{equation}
In a similar way, we find
\begin{equation}
\left(\tilde{m}_\phi^2\right)^\prime\,\simeq\,
2\,m_\phi^2\,a^3\,H
~~~~~\mbox{provided that}~~~~m_\phi\,c^2\,\gg\,10^{-27}\,eV.
\label{eq:24b}
\end{equation}
\begin{equation}
\left(\tilde{m}_\phi^2\right)^{\prime\prime}\,\simeq\,
2m_\phi^2\,a^5H\,\frac{dH}{da} ~~~~~\mbox{provided that}~~~~m_\phi\,c^2\,\gg\,10^{-27}\,eV.
\label{eq:24c}
\end{equation}

Thus, we find that
\begin{equation}
\frac{\tilde{m}_\phi^\prime}{\tilde{m}_\phi^2}\,\ll\,1~~\mbox{and}~~~\frac{\tilde{m}_\phi^{\prime\prime}}{\tilde{m}_\phi^3}\,\ll\,1~~~~~~~~\mbox{provided that}~~~~m_\phi\,c^2\,\gg\,10^{-27}\,eV.
\label{eq:25}
\end{equation}

On the other hand, by Eq.(\ref{eq:5}), we have
\begin{eqnarray}
&&\left|\frac{\omega_k^\prime}{\omega_k^2}\right|\,=\, \left|\frac{\tilde{m}_\phi\tilde{m}_\phi^\prime}{\omega_k^3}\right|\,<\,\left|\frac{\tilde{m}_\phi^\prime}{\omega_k^2}\right|\,<\,
\left|\frac{\tilde{m}_\phi^\prime}{\tilde{m}_\phi^2}\right|, \label{eq:11} \\
&&\left|\frac{\omega_k^{\prime\prime}}{\omega_k^3}\right|\,=\, \left|\frac{1}{\omega_k^4}\left[\tilde{m}_\phi^{\prime\;2}\left(1-\frac{\tilde{m}_\phi^2}{\omega_k^2}\right)
+\tilde{m}_\phi\tilde{m}_\phi^{\prime\prime}\right]\right|
\,<\,\left(\frac{\tilde{m}_\phi^\prime}{\tilde{m}_\phi^2}\right)^2+\left|\frac{\tilde{m}_\phi^{\prime\prime}}{\tilde{m}_\phi^3}\right|. \label{eq:11a}
\end{eqnarray}
Therefore,
\begin{eqnarray}
&&\left|\frac{\omega_k^\prime}{\omega_k^2}\right|\,\ll\,1~~~~ \mbox{provided that}~~~\,\left|\frac{\tilde{m}_\phi^\prime}{\tilde{m}_\phi^2}\right|\,\ll\,1,   \label{eq:12} \\
&&\left|\frac{\omega_k^{\prime\prime}}{\omega_k^3}\right|\,\ll\,1~~~~ \mbox{provided that}~~~\,\left|\frac{\tilde{m}_\phi^\prime}{\tilde{m}_\phi^2}\right|\,\ll\,1~~\mbox{and}
~~\left|\frac{\tilde{m}_\phi^{\prime\prime}}{\tilde{m}_\phi^3}\right|\,\ll\,1.   \label{eq:12a}
\end{eqnarray}

Thus, (\ref{eq:25}), (\ref{eq:12}),
 (\ref{eq:12a}) result in,
\begin{equation}
\left|\frac{\omega_k^\prime}{\omega_k^2}\right|\,\ll\,1
~~~~ \mbox{and}~~~\,
\left|\frac{\omega_k^{\prime\prime}}{\omega_k^3}\right|\,\ll\,1
~~~\,
~~~~ \mbox{provided that}~~~\,  m_\phi\,c^2\,\gg\,10^{-27}\,eV,
 \label{eq:27}
\end{equation}
which are satisfied in $\Lambda$CDM model for $a\,>\,10^{-4}$. Note that the upper limit on the mass of $\phi$ in this equation, namely, $\frac{m_\phi\,c^2}{eV}\,\gg\,10^{-27}$ is satisfied by all standard dark matter candidates including ultra-light dark matter \cite{dark-matter}.

It should be noted that the adiabatic conditions in (\ref{eq:27}) are satisfied independent of the value of $|\vec{k}|$ in $\Lambda$CDM (for $a\,>\,10^{-4}$ and $\frac{m_\phi\,c^2}{eV}\,\gg\,10^{-27}$) \cite{Erdem-Gultekin1,Erdem-Gultekin2}. This is different from the standard form of adibatic conditions that simply impose $\left|\frac{\omega_k^{\prime\prime}}{\omega_k^3}\right|\,\ll\,1$ and $\left|\frac{\omega_k^\prime}{\omega_k^2}\right|\,\ll\,1$ \cite{QFTC2,Boyanovsky} which depend on the value of $|\vec{k}|$. Eq.(\ref{eq:10}) in general is guaranteed only for $|\vec{k}|\,\rightarrow\,\infty$ while otherwise its validity depends on the values of  $|\vec{k}|$, $\tilde{m}_\phi^\prime$, $\tilde{m}_\phi$. On the other hand, (\ref{eq:10}) is satisfied for all values of $|\vec{k}|$ when $\frac{\tilde{m}_\phi^\prime}{\tilde{m}_\phi^2}\,\ll\,1$ and $\left(\frac{\tilde{m}_\phi^{\prime\prime}}{\tilde{m}_\phi^3}\right)\,\ll\,1$ are satisfied, which is the form of the adiabatic conditions in (\ref{eq:27}).

\section{gravitational particle production after decoupling}

\subsection{mode function in adiabatic approximation}

In the view of (\ref{eq:10}) we may take $W_k\,\simeq\,\omega_k$ in (\ref{eq:8}). Therefore, after decoupling
(i.e. after the redshifts of order of $10^4$) we may let \cite{Ford}
\begin{equation}
v_k(\eta)\,=\,\frac{1}{\sqrt{\omega_k(\eta)}}\,\exp{\left[i\int_{\eta_0}^\eta\,\omega_k(\eta)\,d\eta \right]}  \label{eq:31}
\end{equation}
where
\begin{equation}
\omega_k\,\simeq\,\frac{1}{\hbar}\sqrt{\hbar^2\vec{k}^2c^2\,+\,m_\phi^2c^4a^2}\,=\,c\,\sqrt{\vec{k}^2\,+\,\left(\frac{m_\phi\,c}{\hbar}\right)^2\,a^2}.
 \label{eq:44}
\end{equation}
In this section we will also discuss the phenomenological implications of this analysis. To this end, in the following we express the formulas in the forms where $\hbar$ and $c$ are explicitly written i.e. $\hbar$ and $c$ are not set equal to 1.

\subsection{in and out states}

One may take $\omega_k$ approximately consatant in a time interval $\Delta\,\eta$ if
\begin{equation}
\left|\frac{\Delta\,\omega_k}{\omega_k}\right|\,=\,
\left|\frac{\Delta\,\eta\,\left(\frac{d\omega_k}{d\eta}\right)}{\omega_k}\right|\,\ll\,1.
\label{eq:34}
\end{equation}
If (\ref{eq:34}) is satisfied, then (\ref{eq:3}) may be expressed in its Minkowski form in the interval $\Delta\,\eta$, so the field may expanded as \cite{Erdem-Gultekin1,Erdem-Gultekin2}
\begin{eqnarray}
\tilde{\phi}_   {(i)}(\vec{x},\eta)\,\simeq\,
\int\,
\frac{d^3\tilde{p}}{(2\pi)^\frac{3}{2}\sqrt{2\omega_{p,(i)}}}\left[a_{p,(i)}^-\,
e^{i\left(\vec{\tilde{p}}.\vec{r}-\omega_{p,(i)}(\eta-\eta_i)\right)}
\,+\,a_{p,(i)}^+
\,e^{i\left(-\vec{\tilde{p}}.\vec{r}+\omega_{p,(i)}(\eta-\eta_i)\right)}\right]\label{eq:35} \\
\eta_i\,<\,\eta\,<\,\eta_{i+1}\;, \nonumber
\end{eqnarray}
where $_{(i)}$ refers to the $i$th time interval between the times $\eta_i$ and $\eta_{i+1}$ with $\Delta\,\eta\,=\,\eta_{i+1}-\eta_i$. It has been shown in \cite{Erdem-Gultekin2} that (\ref{eq:34}) may be easily imposed for $\Lambda$CDM. Note that (\ref{eq:34}) may be guaranteed by taking $\Delta\,\eta$ sufficiently small once (\ref{eq:27}) is imposed i.e.
\begin{equation}
\frac{\omega_k^\prime}{\omega_k^2}\,=\,\delta\,=\,\frac{\Delta\,\eta\,\omega_k^\prime}{\Delta\,\eta\,\omega_k^2}
\,=\,\left|\frac{\Delta\,\omega_k}{\omega_k}\right|\,\frac{1}{\Delta\eta\,\omega_k}\,\ll\,1
~~~~ \Rightarrow~~~~~~\left|\frac{\Delta\,\omega_k}{\omega_k}\right|\,=\,\delta\,\Delta\eta\,\omega_k\,\ll\,1
 \label{eq:36}
\end{equation}
provided that $\Delta\eta$ is sufficiently small. (However, (\ref{eq:27}) is not guaranteed by (\ref{eq:34})).

In other words, (\ref{eq:34}) is always satisfied provided that $\Delta\eta$ is sufficiently small. For such a $\Delta\eta$ the spacetime essentially may be considered as a Minkowski spacetime. However, $\Delta\eta$ cannot be arbitrarily small. The de Broglie wavelength of the the relevant modes  must be significantly smaller than the size of $c\,\Delta\eta$ so that the detectors at the $in$ and $out$ regions can detect them (as free modes in Minkowski space-time) \cite{Erdem-Gultekin2}. This puts a lower bound on the values of $|\vec{k}|$ (for which this method is applicable) for a given $c\,\Delta\eta$ (and vice versa) through
\begin{equation}
  |\vec{k}|\,>\,|\vec{k}|_\Delta\,=\,\frac{2\pi}{c\,\Delta\eta}.
\label{eq:36a}
\end{equation}

$\Delta\eta\,$s that satisfy (\ref{eq:34}) can also be taken considerably wide in $\Lambda$CDM as shown below.
Eq.(\ref{eq:34}) implies that
\begin{equation}
\Delta\eta\,\ll\,\frac{\omega_k}{\omega_k^\prime}\,=\,\frac{\hbar^2\vec{k}^2\,+\,m_\phi^2c^2a^2(\eta)}{a^3(\eta)\,m_\phi^2c^2H(\eta)}.
 \label{eq:36aa}
\end{equation}
The right-hand side of (\ref{eq:36aa}) is minimum at $|\vec{k}|\,=\,0$, and $H\,\sim\,a^{-\frac{3}{2}}\,H_0$ at the time of decoupling $a_{dec}\,\sim\,10^{-3}$. These imply that $\Delta\eta\,\ll\,\frac{10^{\frac{3}{2}}}{H_0}$ after decoupling and $\Delta\eta\,\ll\,\frac{1}{H_0}$ at present. This, in turn, implies that $\Delta\eta$ values can be taken long enough to identify the  $in$ and $out$ vacuum states of S-matrix formulation in the form of (\ref{eq:35}) \cite{Erdem-Gultekin2}. Note that $\Delta\eta\,\sim\,\frac{10^{\frac{3}{2}}}{H_0}$ at the time of decoupling and  $\Delta\eta\,\sim\,\frac{1}{H_0}$ at present are upper bounds on $\Delta\eta$ for which the spacetime can approximately be taken to be a Minkowski spacetime. As $\Delta\eta$ gets smaller the approximation becomes a better one. However, the price to be paid for a smaller $\Delta\eta$ is that more modes with shorter wavelengths (i.e. with higher $|\vec{k}|$) become excluded from the domain of the applicability of the approximation used in this study. This point will be discussed in the next subsection in more detail.

A mode function corresponding to a ground state with minimum energy at time $\eta_0$ has the form $v_k(\eta_0)\,=\,\frac{1}{\sqrt{\omega_k\left(\eta_0\right))}}\,e^{i\sigma_k\left(\eta_0\right)}$ where $\sigma_k$ is an arbitrary function of $|\vec{k}|$ and $\eta_0$ \cite{QFTC1}. In the light of this and the above observations, the mode functions of out states, for example, may be expressed as mode functions of Minkowski space in each interval $\Delta\eta$ that satisfies (\ref{eq:34}) i.e.
\begin{equation}
v_k^{(out)}(\eta)\,=\,\frac{1}{\sqrt{\omega_k\left(\eta_f\right))}}\,e^{i\left[\omega_k\left(\eta_f\right)\,\eta\right]}  \label{eq:37}
\end{equation}
where
\begin{equation}
\omega_k\left(\eta_f\right)\,\simeq\,c\,\sqrt{\vec{k}^2\,+\,\left(\frac{m_\phi\,c}{\hbar}\right)^2\,a_f^2}.
  \label{eq:37a}
\end{equation}
 is the value of $\omega_k$ at a final time $\eta_f$, and $\eta_f-\Delta\eta\,<\,\eta\,<\,\eta_f$ with $\Delta\eta$ being sufficiently small so that (\ref{eq:34}) holds while not being extremely small. The mode functions of the in states may expressed in the same form as (\ref{eq:37}) where $\eta_f$ is replaced by initial time $\eta_i$, and the in states evolve at later times as in (\ref{eq:31}). (To be precise we identify  $\eta_i$ and $\eta_f$ as $a_i\,=\,a(\eta_i)\,\simeq\;~10^{-4}~-~10^{-3}$, $a_f\,=\,a(\eta_f)\,\simeq\;~1$.)

 \subsection{gravitational particle production}

Now we use the matching conditions for the mode functions and their derivatives at boundaries at the time $\eta$ with $\eta_f-\Delta\eta\,<\,\eta\,<\,\eta_f$, namely,
\begin{eqnarray}
v_k^{(in)}(\eta)&=& \alpha_k\,v_k^{(out)}(\eta)\,+\,\beta_k\,v_k^{(out)*}(\eta)\,  \label{eq:38a} \\
v_k^{(in)\prime}(\eta)&=& \alpha_k\,v_k^{(out)\prime}(\eta)\,+\,\beta_k\,v_k^{(out)*\prime}(\eta)\,  \label{eq:38b}
\end{eqnarray}
to determine $\beta_k$ (since $\left|\beta_k\right|^2$ is the number density of the gravitationally produced particles with momentum $\vec{k}$). Note that $v_k^{(in)}(\eta)$ in \ref{eq:38a}) and (\ref{eq:38b}) is its value in the $out$ region. In this region the form of $v_k^{(in)}(\eta)$ is given by (\ref{eq:31}) while that of $v_k^{(out)}(\eta)$ is given by (\ref{eq:37}).

By (\ref{eq:37}),
\begin{eqnarray}
\alpha_k\,v_k^{(out)}(\eta)\,+\,\beta_k\,v_k^{(out)*}(\eta)\,&=& \frac{1}{\sqrt{\omega_k\left(\eta_f\right))}}\,
\left[\alpha_k\,e^{i\left[\omega_k\left(\eta_f\right)\,\eta\right]}  \,+\,\beta_k\, e^{-i\left[\omega_k\left(\eta_f\right)\,\eta\right]} \,\right]
\label{eq:39a} \\
\alpha_k\,v_k^{(out)\prime}(\eta)\,+\,\beta_k\,v_k^{(out)*\prime}(\eta)\,&=&
i\sqrt{\omega_k\left(\eta_f\right)}\,
\left[\alpha_k\,e^{i\left[\omega_k\left(\eta_f\right)\,\eta\right]}  \,-\,\beta_k\, e^{-i\left[\omega_k\left(\eta_f\right)\,\eta\right]} \,\right]
\label{eq:39b}
\end{eqnarray}
In the following we will let $\eta_f=\eta$ since we consider generic $\eta_f$ i.e. we may replace $\eta_f$ in (\ref{eq:39a}) and (\ref{eq:39b}) by $\eta$ provided that $\eta$ is in a sufficiently small interval $\Delta\eta$.

Hence, after using (\ref{eq:31}) for $v_k^{(in)}(\eta)$ and making use of (\ref{eq:38a}), (\ref{eq:38b}), (\ref{eq:39a}), (\ref{eq:39b}), we find
\begin{equation}
\left|\beta_k\right|^2
\,=\,\frac{\left(\frac{m_\phi\,c}{\hbar}\right)^4\,a^2\,a^{\prime\,2}}{16c^2\left[\vec{k}^2\,+\,\left(\frac{m_\phi\,c}{\hbar}\right)^2a^2\right]^3}
\label{eq:47}
\end{equation}

Thus,
\begin{equation}
\bar{n}\,=\,\frac{1}{(2\pi)^3}\i\int\,d^3k\,\left|\beta_k\right|^2
\,=\,\frac{1}{512\,\pi}\left(\frac{m_\phi\,c}{\hbar}\right)\,a^3\,\left(\frac{H}{c}\right)^2
\label{eq:48}
\end{equation}
where $\bar{n}$ is the number density of gravitationally produced particles in the comoving coordinates. Note that $\vec{k}$ is the wave number vector in comoving coordinates (rather than the physical wave number vector $\frac{1}{a}\vec{k}$). Hence, the physical number density is
\begin{eqnarray}
&&n\,=\,\frac{\bar{n}}{a^3}\,=\,\frac{1}{512\,\pi}\left(\frac{m_\phi\,c}{\hbar}\right)\,\left(\frac{H}{c}\right)^2\label{eq:48a}
\end{eqnarray}

\subsection{energy density of gravitationally produced particles}

The energy density corresponding to (\ref{eq:47}) is
\begin{equation}
\rho^{(PP)}\,=\, \frac{1}{(2\pi)^3\,a^3}\int\,d^3k\,E_k\,\left|\beta_k\right|^2
\,=\,\frac{\hbar}{96\,\pi\,c}\,
\left(\frac{m_\phi\,c}{\hbar}\right)^2H^2
\label{eq:48}
\end{equation}
where $E_k\,=\,\sqrt{\hbar^2\left(\frac{\vec{k}}{a}\right)^2c^2\,+\,m_\phi^2c^4}$ is the physical energy of a $\phi$ particle with physical momentum $\frac{1}{a}\hbar\vec{k}$ (while $\hbar\vec{k}$ is the comoving coordinate momentum of the particle).

The total effective energy density $\rho^{eff}$ (that is the sum of the energy density of background $\rho^{(bg)}$ and the energy density of gravitationally produced particles $\rho^{(PP)}$) relates to Hubble parameter $H$ as
\begin{eqnarray}
\frac{3\,c^2}{8\pi\,G}\,H^2&&=\,\rho^{eff}\,=\,\rho^{(PP)}\,+\,\rho^{(bg)}
\,=\,\frac{\hbar}{96\,\pi\,c}\,\sum_i\left(\frac{m_i\,c}{\hbar}\right)^2\,H^2
\,+\,\rho^{(bg)}
\nonumber\\
&\simeq&\rho^{(bg)}
\,+\,1.8\,\times\,10^{-58}\times\,
\sum_i\left(\frac{m_i\,c^2}{eV}\right)^2\,
\frac{3\,c^2}{8\pi\,G}\,H^2
\label{eq:49c} \\
&&\mbox{i.e.}~~~3\,H^2\,=\,\frac{8\pi}{c^2}\,\left[\frac{G}{1\,-\,1.8\,\times\,10^{-58}\times\,
\sum_i\left(\frac{m_i\,c^2}{eV}\right)^2}\right]\,\rho^{(bg)} \label{eq:49d}
\end{eqnarray}
where $\rho^{(bg)}$ is identified as the total energy density, and $m_i$ denotes the mass of the $ith$ scalar particle (that contributes to $\rho^{(PP)}$). Eq.(\ref{eq:49d}) implies that gravitational particle production has a significant contribution to the effective Hubble parameter if $\sum_i\left(\frac{m_i\,c^2}{eV}\right)^2$ is not extremely smaller than $10^{58}$. For example, if there are ten (scalar) particles with masses of the order of the the the Planck mass $M_{Planck}\,\simeq\,1.22\times\,10^{28}\,eV/c^2$ (that were, for example, present at the beginning of the universe and later may had decayed wholly into standard model particles) then $G$ would be multiplied by an overall factor $\sim$$\,1.37$ in (\ref{eq:49d}). Note that, ultra-heavy particles are extensively studied in literature \cite{heavy-DM,Chung,Grib}.

A comment is in order here. The present study is applicable for the times after the time of decoupling till the present.  We have checked the applicability of Eq.(\ref{eq:27}) in this interval (which is in the order of $\frac{1}{H_0}$). In a similar way we have found in Section III B. that identification of the $in$ state in an unambiguous way requires $\Delta\eta$ to be smaller than $\frac{\sqrt{10^{3}}}{H_0}$, and identification of the $out$ state in an unambiguous way requires $\Delta\eta$ to be smaller than $\frac{1}{H_0}$. Therefore, the present analysis is valid for wavelengths smaller than
$\frac{1}{H_0}$.  This corresponds to a lower bound on the relevant modes, namely, $\hbar\,|\vec{k}|\,c\,\sim\,10^{-33}\,eV$. For modes with lower $|\vec{k}|$, the validity of the approximation cannot be guaranteed. However, the contribution of such low momentum modes in Eq.(\ref{eq:48}) is small (unless there is a drastic change in the form of $\left|\beta_k\right|^2$ in Eq.(\ref{eq:47}) at such low values of $|\vec{k}|$). Therefore, Eq.(\ref{eq:48a}) and Eq.(\ref{eq:49d}) may be taken as decent approximate expressions. In fact, the same formulas Eq.(\ref{eq:48a}) and Eq.(\ref{eq:48}) are obtained in literature \cite{Starobinsky,book,Grib2}.

It is evident from (\ref{eq:49d}) that gravitational production of $\phi$ particles results in an effective overall increase in the value of Hubble parameter, hence in the value of Hubble constant. This increase, at first sight, may be attributed either to an effective increase in the Newton's gravitational constant $G$ or to an effective increase in the total energy density. However, such an effective increase in the total energy density cannot be considered to be due to a physical increase in the energy density of background particles (e.g. baryons). Increasing the mass of $\phi$ results in an overall increase in the total energy density irrespective of the masses and the ratio of the particles in the background. Gravitational particle production is not specific to scalars. It is possible for all particles \cite{book,Chung2,vector} but their contribution to total energy density is proportional to their masses in all cases. Hence, if the total mass of the scalar particles are taken to be very large e.g. at order Planck mass while all other particle masses are taken to be much smaller, then the increase in total energy density will be determined by the total mass of the scalars. In such a case the effective total energy density increases significantly while the energy density of the background particles such as baryons virtually remain the same. This point will be important in the discussion in the paragraph after Eq.(\ref{eq:58}) (i.e. in the argument that the number density of baryons essentially remain the same in such a case while the effective Hubble constant in direct measurements increases considerably). Moreover, the effective increase in the value of the Hubble parameter cannot be also attributed to a true physical production of $\phi$ particles since a true physical production of scalar particles would induce an energy density for a scalar field in Eq.(\ref{eq:49d}) that scales as that of a scalar field. (Note that, in principle, the energy density of a scalar field may mimic energy density of any fluid e.g. of $\Lambda$CDM while it cannot exactly be the same as that of that fluid for a finite time.) On contrary, there is no energy density that scales as that of a scalar field in Eq.(\ref{eq:49d}) if $\rho^{(bg)}$ is taken as the energy density of $\Lambda$CDM). This point i.e.
$\rho^{(PP)}$ above should be identified as the effective energy density due to quasi-particles \cite{book} rather than true particles can be also seen in the following way. Identification of $\rho^{(PP)}$ as (effective) energy density of true physical particles would lead to an inconsistency. If $\rho^{(PP)}$ were a true energy density it would increase the total energy density, hence increase $H$, this, in turn, would induce additional gravitational production of particles, this, in turn, would increase the total energy density further, and eventually the total energy density would be infinite. In other words such an argument would eventually result in $\rho^{(PP)}\,\propto\,lim_{N\rightarrow\,\infty}\,(1-\gamma)^{-N}\,\rightarrow\,\infty$ where
$\gamma= 1.8\,\times\,10^{-58}\times\,\left(\frac{m_\phi\,c^2}{eV}\right)^2$. In the light of the above consideration it is more conceivable and reliable to identify the effective increase in the Hubble parameter and the Hubble constant to be due to an effective increase in $G$ as described in (\ref{eq:49d}).  This effect may be significant, for example, for scalar particles that were present at the extremely early times (e.g. at the time of inflation and then decayed wholly into other particles) with masses at the order of Planck masses. Another comment is that gravitational particle production does not modify the evolution of energy density as is evident in (\ref{eq:49d}) (since the effect of gravitational particle production is to multiply the background energy density by an overall constant as is evident in Eq.(\ref{eq:49d}).  On the other hand, a true physical production of scalar particles would induce an energy density that scales as that of a scalar field. Therefore, as  mentioned above, the effective energy density induced by gravitational particle production of $\phi$ particles in the present context should be identified as the energy density of quasi-particles rather than that of physical particles. In fact it is possible to consider the case where there is also a contribution to the energy density by physical $\phi$ particles. In that case there would also be a contribution to the Hubble parameter that scales as that of a scalar field. All these imply that it is more appropriate to identify the overall effective increase in the Hubble parameter (due to gravitational particle production) to be an effective increase in $G$ as in (\ref{eq:49d}) rather than an increase in the energy density. The gravitational particle production in this paper involves $\phi$ particles that are not physically produced and the effective value of the gravitational constant is increased by gravitational particle production. This mechanism is analogous to vacuum polarization in quantum electrodynamics (QED). Vacuum polarization in QED (after renormalization) causes an effective re-scaling in the electromagnetic coupling constant due to pairs of electrically charged virtual pairs rather than physical produced particles.  Therefore, the results obtained in the present study may be considered to be due to some sort of gravitational vacuum polarization \cite{Starobinsky, QFTC2}.

\section{Impact of gravitational particle production on the Hubble tension}

The effective increase of $G$ in (\ref{eq:49d}) causes an increase in the overall value of the Hubble parameter, so an increase in Hubble constant, namely,
 \begin{equation}
H_0^2\,=\,\left[\frac{1}{1\,-\,1.8\,\times\,10^{-58}\times\,
\sum_i\left(\frac{m_i\,c^2}{eV}\right)^2}\right]\,\bar{H}_0^2 \;. \label{eq:49da}
\end{equation}
where the subscript $0$ stands for the present time, and $\bar{H}_0\,=\,\sqrt{\frac{8\pi\,G}{3}\,\rho_0}$ is the value of the Hubble constant without the effect of gravitational particle production included while $H_0$ is the value of the Hubble constant after inclusion of the effect of gravitational particle production. Note that, by (\ref{eq:49d}), $H_0$ is the value of the Hubble constant determined in direct measurements.

 Hubble constant may be also determined from the imprints of baryon acoustic oscillations on CMB or large scale structure anisotropies by measuring the angle $\theta$ subtended by sound horizon
\begin{equation}
\theta\,=\,\frac{r_s}{D_A} \label{eq:54}
\end{equation}
where $r_s$ is the comoving size of the sound horizon, $D_A$ is the comoving angular diameter distance to the observed position. Here \cite{Kamionkowski,BAO}
\begin{equation}
r_s\,=\,\int_{z_a}^\infty\,\frac{v_s(z)\,dz}{H_0\,E(z)}~,~~~~
D_A\,=\,c\int_0^{z_b}\,\frac{dz}{H_0\,E(z)}
\label{eq:54a}
\end{equation}
where  $z$ denotes redshift; $c$ is the speed of light; $v_s(z)$ is the speed of the sound waves in baryon-photon fluid; $a$ = * or $d$ stand for recombination or drag epoch (for the imprint of the acoustic oscillations on CMB radiation or on galaxy autocorrelation function, repectively); $b$ = * or $obs$ denote the redshifts of recombination or of the observed galaxies; $E(z)\,=\,\sqrt{\Omega_\Lambda\,+\,\Omega_M\,\left(1+z\right)^3\,+\,\Omega_R\,\left(1+z\right)^4}$ in $\Lambda$CDM with $\Omega_\Lambda$, $\Omega_M$, $\Omega_R$ being the density parameters for cosmological constant, dust, radiation, respectively.

Let us assume (unlike the early or late time solutions of the Hubble tension) that the evolution of the universe before and after the recombination are described by the (unmodified) standard model (i.e. $\Lambda$CDM). (In fact, we have expressed (\ref{eq:54a}) in a form that is more suitable for this case.) One observes that $\theta$ in (\ref{eq:54}) is unaffected by the values of $H_0$ in the arguments of $r_s$ and $D_A$. However, the value of the Hubble constant affects $r_s$ and $D_A$ by its effect on $z_a$ by affecting recombination as we will see below. The effects of a change in $z_a$ on $r_s$ and $D_A$ are not the same since the value of $r_s$ is dominated by the value of $E(z)$ at values of $z$ close to $z_a$ while the value of $D_A$ is dominated by the value of $E(z)$ at values of $z$ close to $z=0$. Hence, a variation of the Hubble constant varies $\theta$ by its effect on $z_a$. Thus, the observational value of the Hubble constant may be determined after finding the best fit values for the Hubble constant and the density parameters corresponding to the observed $\theta$. Below, we will see that the Hubble constant determined in this way is its value without the contribution of gravitational particle production i.e. $\bar{H}_0$ (while the value of the Hubble constant that is determined in direct measurements is $H_0$). First we will present the argument in the context of Saha equation to see the situation in an easier way at conceptual level. Then, we will reconsider the situation at the level of the corresponding Boltzmann equation to obtain essentially the same result in more concrete terms in a more rigorous way.

The general aspects of recombination may be studied by Saha equation \cite{Weinberg}
\begin{equation}
X\,(1\,+\,S\,X)\,=\,1  \label{eq:55}
\end{equation}
where $X\,=\,\frac{n_p}{n_p+n_{1s}}\,=\,\frac{n_e}{n_p+n_{1s}}$ is the fraction of protons or electrons to the total number of baryons (i.e. protons plus neutral hydrogen atom\textcolor{blue}{s}), and
\begin{equation}
S\,=\,0.76\,n_b\,\left(\frac{m_ek_B\,T}{2\pi\hbar^2}\right)^{-\frac{3}{2}}\,\exp{\frac{B_1}{k_BT}}.  \label{eq:56}
\end{equation}
Here $n_b$, $m_e$, $k_B$, $B_1$ are the number density of baryons (at temperature T), electron mass, Boltzmann constant, the binding energy of hydrogen atom in its ground state; respectively. The decoupling of photons from baryons took place at a sufficiently small value of $X$, say at $X_*\,\ll\,1$. It is evident from (\ref{eq:55}) that the value of $X$ is determined by the value of $S$ which is related to $n_b$ by (\ref{eq:56}).  $n_b$ is related to the number density at present time $n_{b0}$ by $n_b\,=\,n_{b0}\,\left(\frac{T}{T_{\gamma0}}\right)^3$ where $T_{\gamma0}\simeq\,2.73\,K$ is the present day temperature of CMB. $n_{b0}$ is calculated by using
\begin{equation}
n_{b0}\,=\,\frac{3\Omega_b\,H_0^2}{8\pi\,G^{(effective)}\,m_N}\,=\,\frac{3\Omega_b\,\bar{H}_0^2}{8\pi\,G\,m_N}
\,=\,1.121\times\,10^{-5}\,\Omega_b\,\bar{h}^2\;\mbox{nucleons}/cm^3 \label{eq:57}
\end{equation}
where $\Omega_b$ is the density parameter for baryons.
Here, essentially (\ref{eq:49d}) is used where $G$ in \cite{Weinberg} is replaced by its effective value $G^{(effective)}$, and $H_0$ is identified as the effective value of the Hubble constant in Friedmann equation (i.e. in (\ref{eq:49d})) that includes the contribution due to gravitational particle production. $\bar{H}_0$ is the value of the Hubble constant before inclusion of the effect of gravitational particle production, and $\bar{h}\,=\,\frac{\bar{H}_0}{100~km\;s^{-1}\;Mpc^{-1}}$. It is evident from  (\ref{eq:57}) that the parameter that determines the evolution of the photon-baryon plasma before decoupling is $\bar{h}$ rather than $h$.

Although, Saha equation is enough to give the basic elements of the evolution the photon-baryon plasma it has some important shortcomings. The first shortcoming is that it does not specify the exact value of $z_*$. The second  is that Saha equation is derived by assuming chemical equilibrium in the scattering $e^-$ + p $\leftrightarrow$ {\bf H} + $\gamma$ (where {\bf H} denotes hydrogen atom) while chemical equilibrium is not applicable at the time of decoupling. Finally, Saha equation describes the evolution of the background while CMB anisotropies and BAO calculations are at the level of cosmological perturbations. These shortcomings may be removed by using the Boltzmann equation corresponding to this case. The photon-baryon system at the time of recombination has kinetic equilibrium (while not necessarily chemical equilibrium) and the electron\textcolor{blue}{s} are non-relativistic. The corresponding Boltzmann equation is \cite{Dodelson}
\begin{equation}
\frac{dX}{dt}\,=\,\left[<\sigma\,v>\left(\frac{m_ek_B\,T}{2\pi\hbar^2}\right)^{\frac{3}{2}}\,\left(1\,-\,X\right)\exp{\{-\left(m_e+m_p-m_H\right)c^2/(k_BT)\}}
\,-\,<\sigma\,v>\,n_b\,X^2\right]
\label{eq:58}
\end{equation}
where $n_e<\sigma\,v>$ is thermally averaged rate for the decrease of electrons in $e^-$ + p $\leftrightarrow$ H + $\gamma$. Note that $n_b$ in (\ref{eq:58}) is related to $n_{b0}$ in (\ref{eq:57}) (that depends on $\bar{h}$ rather than $h$). The equation (\ref{eq:58}) may be integrated numerically to have a detailed evolution of $X$, and $z_*$ (for given values of $\bar{h}$ and the density parameters). $z_*$ may be determined by finding the value of $z$ where there is sharp decrease in $X$ i.e. by finding  $X_*\,\ll\,1$ where $X$ drops sharply. Hence the best fit values of $\bar{h}$ and the density parameters may be determined by using Boltzmann codes such as CAMB \cite{Planck}. In fact, this is how the Hubble constant is determined in CMB and BAO calculations. One may get further insight into the problem by analytic formulas that expresses $z_*$ and $z_d$ in termd of $\bar{h}^2\Omega_M$ and $\bar{h}^2\Omega_b$ \cite{Hu1}
\begin{equation}
z_*\,=\,1048\,\left[1+0.00124\,\left(\Omega_b\bar{h}^2\right)^{-0.738}\right]\left[1+g_1\left(\Omega_M\bar{h}^2\right)^{g_2}\right]
\label{eq:58a}
\end{equation}
\begin{equation}
z_d\,=\,1315\,\frac{\left(\Omega_M\bar{h}^2\right)^{0.251}}{1\,+\,0.659\,
\left(\Omega_M\bar{h}^2\right)^{0.828}}\left[1+b_1\left(\Omega_b\bar{h}^2\right)^{b_2}\right]
\label{eq:58b}
\end{equation}
where $h$ in \cite{Hu1} is replaced by $\bar{h}$ (since the dependence of (\ref{eq:58}) on the Hubble constant is through $n_b$ which is unaffected by gravitational production of $\phi\,$s). Here, $g_1$, $g_2$ are some functions of $\Omega_b\bar{h}^2$ and $g_1$, $g_2$ are some functions of $\Omega_M\bar{h}^2$ whose explicit forms may be found in \cite{Hu1}. The effect of $n_b$ on $z_*$ and $z_d$ (through its dependence on $\Omega_b$) is evident in (\ref{eq:58a}) and (\ref{eq:58b}). Note that (\ref{eq:58a}) and (\ref{eq:58b}) are functions of  $\Omega_M\bar{h}^2$ and $\Omega_b\bar{h}^2$ rather than being functions of $\Omega_M$, $\Omega_b$, $\bar{h}$. $D_A$ in (\ref{eq:54a}) may be expressed in terms of $\Omega_M\bar{h}^2$ and $\Omega_\Lambda\bar{h}^2$ (where the contribution of radiation may be neglected since value of $D_A$ is dominated by low redshift contributions) and $r_s$ in (\ref{eq:54a}) may be expressed in terms of $\Omega_M\bar{h}^2$ and $\Omega_r\bar{h}^2$ (where the contribution of cosmological constant may be neglected since value of $r_s$ is dominated by the redshifts close to $z_*$). As we have remarked in the discussion after (\ref{eq:54a}), although the Hubble constants in $H(z)$ of $D_A$ and $r_s$ cancel in (\ref{eq:54}), $z_*$ remains dependent on $\Omega_M\bar{h}^2$. Therefore, we may express $D_A$ and $r_s$ in terms of the the density parameters times $\bar{h}^2$. This implies that what we obtain through data fit for $\theta$ are $\Omega_M\bar{h}^2$, $\Omega_b\bar{h}^2$. Therefore, by the observation of $\theta$ one cannot obtain the value of $\bar{h}$ separately. However, one may use a phenomenological rule observed by \cite{Percival}, namely, in a spatially flat universe $\Omega_M\bar{h}^{p}$ (where $p=3.4$ in the original paper while $p$ is found to be 3 by Planck) may be determined from the positions of the accoustic peaks (while $\Omega_M\bar{h}^2$ may be directly determined from data analysis for best fists.). This information may be used to determine $\Omega_M$, $\bar{h}$ (and $\Omega_b$) separately \cite{Planck,Percival}.

To summarize, $\bar{H}_0$ is the value obtained by Planck \cite{Planck} (for the Planck data set) and does not contain a contribution from gravitational particle production (GPP) while $H_0$ is the directly measured value of the Hubble constant that has contribution from GPP. The difference between $H_0$ and $\bar{H}_0$ may be wholly attributed to GPP if the value of  $\sum_i\left(\frac{m_i\,c^2}{eV}\right)$ is taken accordingly. In any case, GPP ameliorates the Hubble tension. It should be remarked that no new physics is employed in the present study. The standard $\Lambda$CDM model (without any extension) is employed here. The only difference between this study and the other studies in the past that employed $\Lambda$CDM model is the inclusion of GPP that is neglected in the other studies. What has been done here is not introducing a new model. What has been done here is to give an explanation for observing two different values of the Hubble constant in direct and indirect measurements. It has been shown here that GPP modifies the directly measured value of the Hubble constant $H_0$ while it leaves the value of the Hubble constant in CMB measurements $\bar{H}_0$ intact. $\bar{H}_0$ is obtained from number density of baryons $n_b$ that is unaffected by gravitational production (as seen in (\ref{eq:57})) while $H_0$ is obtained from Eq.(\ref{eq:49d}) which includes the effect of GPP. No new model is introduced in this paper. The model employed here is just the standard $\Lambda$CDM model (where the effect of GPP is included). The effect of the GPP, as is evident from Eq.(\ref{eq:49d}), is multiplying the Hubble parameter of the background by an overall constant. Therefore, no new data analysis (in addition to that of $\Lambda$CDM) is needed for CMB and BAO data sets (unlike the extensions of $\Lambda$CDM model \cite{why}). The values obtained from these data sets (with $\Lambda$CDM adopted) remain applicable here. The point here is that the values of the Hubble constant obtained by the use of the CMB and BAO anisotropy data versus the corresponding formula (\ref{eq:54}) and (\ref{eq:54a}) are employed for the best fit value of $z_*$ or $z_d$ which in turn are determined by $n_b$, so by $\bar{H}_0$.  Hence, $\bar{H}_0$ corresponds to the values of the Hubble constant obtained in CMB and BAO observations.

In the second paragraph after Eq.(\ref{eq:49d}) the effective increase in the Hubble constant is identified as an effective increase in the Newton's gravitational constant $G$, rather than an effective increase in the total energy density. It should be remarked that the approach to Hubble tension in the present study is quite different from the models with a jump in the value of $G$ at very small redshifts \cite{jump-G1, jump-G2}. Those type of models need a rigorous theoretical motivation and  do not solve the Hubble tension wholly (while they ameliorate it) \cite{jump-G3}, and data seems not to support the prediction of those models that $H_0$ should vary when obtained in different redshift bins \cite{jump-G4}. The gravitational constant $G$ in those studies varies with redshift while the gravitational constant in the present study does not vary with redshift. Moreover, the model we employ is the standard model of cosmology $\Lambda$CDM and no new physics is used. Only, the effect of gravitational particle production (that is an element of the standard established physics which is overlooked in the previous studies) is taken into account. The inclusion of this effect explains why the value of the Hubble constant in the direct measurements and in the CMB and BAO calculations are different. No need for additional numerical simulations are needed. What is done is just the usual $\Lambda$CDM data analysis that have been done by CMB and BAO collaborations. In other words, what is done in this paper is to give an explanation for having two different values of the Hubble constant obtained from  direct measurements and CMB and BAO collaborations rather than proposing a new model. The model employed in this paper is just $\Lambda$CDM (both at the background and at the level of perturbations) since it just amounts to multiplying the Newton constant by an overall constant as is evident from Eq.(\ref{eq:49d}). This is also different from the case in some models (such as dark energy dark matter coupling models) where the evolution of the background is the same \cite{Erdem} or almost the same as the one in $\Lambda$CDM \cite{Uzan} while their predictions differ at the level of the evolution of cosmological perturbations \cite{Pan}. Instead, the evolution of the Hubble parameter before and after inclusion of the effect of gravitational particle production is the same as that of $\Lambda$CDM.

\section{Conclusion}

Gravitational particle production (GPP) of scalar particles and its contribution to Hubble parameter are studied in the era after decoupling till the present, and their phenomenological implications in the context of the Hubble tension are discussed. No new physics is employed in the present study. The model used here is just the standard $\Lambda$CDM. The only new element is the inclusion of the effect of GPP  that was neglected in previous studies. It is observed that the effect of GPP is to raise the value of the Hubble constant in direct measurements. This effect may be significant if production of extremely heavy scalar particles are allowed phenomenologically at sufficiently high energies (even when they do not exist at present). The raised value of the Hubble constant (due to GPP) is the value of the Hubble constant that is measured in direct local measurements such as the Type Ia supernovae measurements. On the other hand, the value of the Hubble constant relevant to recombination calculations is the one without the effect of GPP. The Hubble parameter after inclusion of GPP is an overall constant times the Hubble parameter before inclusion of GPP. Therefore, the evolution of the Hubble parameter after inclusion of the effect of GPP is the same as its form before inclusion of the effect of GPP. In other words, no new physics is introduced here, only an explanation for the presence of two different classes of measurements of the Hubble constant (from direct and indirect measurements) is given. This may be a clue towards the solution of the Hubble tension. In future, further studies on this topic may be helpful to see all implications and details of the scheme introduced here. In particular, study of possible limitations of the use of gravitational particle production in cosmology (in the view that gravity is studied in a classical setting while matter particles and forces are studied are quantized) may be useful.

\end{document}